\documentclass[11pt]{article}
\usepackage{amsmath}
\usepackage{jheppub}
\usepackage{amssymb}
\usepackage{graphicx}
\usepackage[T1]{fontenc}
\usepackage{slashed}
\usepackage{hyperref}
\usepackage[utf8]{inputenc}
\usepackage{xspace}
\usepackage{color}
\usepackage{ulem}
\usepackage{array}
\usepackage{ulem,fancyvrb}
\usepackage{xcolor}
\usepackage{mathtools}

\newcommand{\abs}[1]{|#1|}

\newcommand{\diag}{\operatorname{diag}}
\allowdisplaybreaks

\date{\today}

\begin{document}

\title{Non-minimal Coupling Inflation and Dark Matter under the $\mathbb{Z}_{3}$ Symmetry}

\author[a,c]{Wei Cheng,}
\affiliation[a]{School of Science, Chongqing University of Posts and Telecommunications, Chongqing 400065, China}
\emailAdd{chengwei@cqupt.edu.cn}

\author[b,1]{Xuewen Liu,}
\affiliation[b]{Department of Physics, Yantai University, Yantai 264005, China}
\emailAdd{xuewenliu@ytu.edu.cn}

\author[a,c,1]{and Ruiyu Zhou}
\affiliation[c]{Department of Physics and Chongqing Key Laboratory for Strongly Coupled Physics, Chongqing University, Chongqing 401331, China}
\emailAdd{zhoury@cqupt.edu.cn}
\footnotetext[1]{Corresponding authors.}

\abstract{We study the cosmological inflation and dark matter (DM) in a unified way within a $Z_3$ complex scalar model. The real and imaginary parts of the complex scalar act as the inflaton and DM respectively. The slow-rolling inflation with non-minimal coupling in both the metric and Palatini formalisms can be realized.
We examine the whole parameters space by fully considering the theoretical and experimental constraints.
We find that in the low-energy scale,
the DM relic density and the DM-nucleon direct scattering experiments favor the mixing angle $|\theta| \lesssim 0.25$, the DM mass $m_\chi \gtrsim 80\rm{GeV}$, and the mass of Higgs-like scalar $m_{h_2} \gtrsim 300\rm{GeV}$.
In the high-energy scale, after further considering the cosmological constraints of the scalar spectral index and the tensor-to-scalar ratio for the two forms of inflation,
the scalar spectral indices are both $\sim 0.965$, the non-minimum coupling coefficients are $\sim 10^4$ and $\sim 10^9$, and the tensor-to-scalar ratios are $\sim 10^{-3}$ and $\lesssim 10^{-11}$ respectively, which suggests that the inflation under the two formalisms can be distinguished by measuring the tensor-to-scalar ratio with higher precision.
}
\maketitle
\preprint{}

\section{Introduction}

Inflation and dark matter (DM), as two major problems in physics, have long puzzled physicists.
Inflation is an early stage of the Big Bang, which is characterized by an extremely small duration of time, and the volume of the universe has been enlarged by a great multiple. This feature allows inflation to gracefully explain the three major problems of the Big Bang model: the horizon problem, the flatness problem, and the magnetic monopole problem ~\cite{Riotto:2002yw}.
As the universe evolves, nowadays, DM has been verified to be an important constituent of the universe by extensive cosmological observations~\cite{Deng:2020dnf}. The missing matter as particles shall be explained by physics beyond the Standard Model (SM).
With the in-depth study of cosmological inflation and DM by theorists and experimentalists, the two parts are increasingly intertwined, which promotes the joint research of inflation and DM~\cite{Cardenas:2007xh,Daido:2017wwb,Choubey:2017hsq,Takahashi:2019qmh,Cheng:2021qmc}.

Numerous studies on inflation and DM have been carried out in the literature. One economic scheme is to treat inflaton and DM as the same particle~\cite{Lebedev:2021zdh,Almeida:2018oid,Borah:2018rca}, which is originally proposed in Refs.~\cite{Liddle:2006qz,Liddle:2008bm}. Ref.~\cite{Hooper:2018buz} discussed the possibility of Weakly Interacting Massive Particle (WIMP) acting as both inflaton and DM, and recent studies \cite{Daido:2017wwb,Cheng:2021qmc} investigated the possibility of axion-like particle acting as both inflaton and DM.
Another common scheme is to describe inflation and DM in one framework, for instance, an inflationary model involving a gauge singlet scalar and fermionic DM was investigated in Ref.~\cite{Aravind:2015xst}, the connections between inflationary observables and DM quantities in the presence of Higgs-portal interactions as well as the nonminimal couplings to gravity was investigated in Ref.\cite{Kim:2014kok}, and a Higgs inflation scenario with a right-handed neutrino and singlet scalar DM was discussed in Ref.~\cite{Haba:2014zda}, while the Higgs inflation extension with two Higgs doublets coupled to the gravity was considered in Ref.~\cite{Gong:2012ri}.
For the former scheme, it’s difficult to consider inflation and DM as the same particles in most proposed models, while the later scheme is more general.

However, the parameter space of the latter scheme in the polynomial-type cosmological inflation potential model is further compressed by the latest Planck observational experimental data~\cite{BICEP:2021xfz}. The usual improvement approach is to consider a non-zero coupling between the Ricci scalar and the matter fields which is called non-minimum coupling correction. This will bring it back to the feasible parameter space~\cite{Kanemura:2012ha,Hashimoto:2020xoz,Rubio:2018ogq}.
There are two ways to deal with the non-minimum coupling~\cite{Bauer:2008zj,Sotiriou:2008rp,Dioguardi:2021fmr,Kodama:2021yrm}, one is metric formalism and the other is Palatini formalism. The main difference between them is that the relationship between connection and metric is different. For the Palatini formalism, the connection and the metric are completely independent, while it is the opposite for the metric formalism, which will lead to different predictions.

The inflation and DM have been extensively discussed in the metric form. For example, Ref.~\cite{Choi:2020ara} studied inflation and DM in the gauged $\mathbb{Z}_{3}$ model. Refs.~\cite{Cheng:2018axr,Cheng:2018ajh} discussed cosmological inflation, phase transitions and DM under the scalar DM model. Palatini formalism is more concerned by cosmologists, such as studies in Refs.~\cite{Amarzguioui:2005zq,Cao:2017jsi,Nojiri:2017ncd,Xu:2017qxf,Odintsov:2020iui,Bekov:2020dww,Zhou:2022ovp,Pan:2021tpk}, however, the non-minimum coupling correction in the Palatini form has less joint study with DM.

In this context, we will use the complex scalar model under $\mathbb{Z}_{3}$ symmetry to study cosmological inflation with both types of non-minimum coupling and DM phenomenology.
Compared with WIMPs under the $\mathbb{Z}_{2}$ symmetry, $\mathbb{Z}_{3}$ symmetry can not only implement the Strongly Interacting Massive Particles (SIMPs) paradigm~\cite{Hochberg:2014dra,Hochberg:2014kqa,Lee:2015gsa},
but also bring lots of interesting signatures at astrophysical observations and intensity frontiers.
We connect the low energy scale physics (DM) to the high energy scale (inflation) by considering the evolution of the renormalization group equation (RGE), which leads to a consistent study of the early universe in a single framework.
To fully explore the parameter range involved in the model, we first scatter the coupling coefficients and impose various  theoretical constraints (perturbation, unitarity, and stability) on the couplings.
We require the theoretical constraints to be satisfied from the electroweak scale up to Planck scale.
Interestingly, this excludes the vast majority of the parameter space.
On this basis, the DM relic density and the direct detection limits are studied,
and the relationship between DM mass and the scalar mixing angle is derived.
Also, we explore the influence of the quartic coupling on inflation, which also involves with DM physics.
Finally, the predictions of the cosmological inflationary observations (scalar spectral index $n_s$, the tensor-to-scalar ratio $r$) under the metric and Palatini forms are analyzed and compared, as well as the magnitude of the corresponding non-minimum coupling coefficient.

This paper is organized as follows.
In Sec.~\ref{sec:Mod}, we introduce the complex singlet model under $\mathbb{Z}_{3}$ symmetry with a non-minimal coupling.
In Sec.~\ref{sec:Inf}, the inflation in the metric and Palatini formalisms are investigated in detail.
In Sec.~\ref{sec:DM}, the relationship between the potential parameters and the physical quantities of DM is investigated.
In Sec.~\ref{sec:Con}, various theoretical and experimental constraints are applied to constrain the parameters of the model. Within these constraints, the numerical analysis of DM relic density, the tensor-to-scalar ratio, and the scalar spectral index are discussed in Sec.~\ref{sec:Num}.
Finally, we will briefly summarize in Sec.~\ref{sec:Sum}.

\section{Model}\label{sec:Mod}

The most general scalar action of the Higgs doublet $H$ and the complex singlet $S$ that couple non-minimally to the Ricci scalar $R$ in the Jordan frame can be written as:
\begin{align}
S_J = \int d^{4}x \sqrt{-g_{J}} \left[ - \frac{M_{\rm Pl}^{2}}{2} R - \frac{F(H,S)}{2} R + D_{\mu}H D^{\mu}H^\dagger + \partial_{\mu}S \partial^\mu S^\dagger - V(H,S) \right],
\end{align}
where $M_{\rm Pl}$ is the reduced Planck mass, $D_\mu$ is the covariant derivative, $F(H,S) =  \alpha H^2 + \beta S^2$ with $\alpha, \beta$ being dimensionless couplings between $H, S$ and gravity respectively. $V(H,S)$ is the invariant scalar potential with a $\mathbb{Z}_3$ symmetry, i.e., $H \to H$, $S \to e^{i 2 \pi/3} S$, which can be concretized as
\begin{equation}
\begin{split}
  V(H, S) &= \mu_{\rm H}^{2} \abs{H}^{2} + \lambda_{H} \abs{H}^{4}
  + \mu_{S}^{2} \abs{S}^{2} + \lambda_{S} \abs{S}^{4}
   \\
  & + \lambda_{SH} \abs{S}^{2} \abs{H}^{2} + \frac{\mu_3}{2} (S^{3} + S^{\dagger 3}).
\end{split}
\label{eq:V:Z:3:singlet}
\end{equation}

Note that any quartic coupling terms have been excluded under $\mathbb{Z}_{3}$ symmetry, and the cubic $\mu_{3}$ term breaks a $\mathbb{Z}_3$ symmetry softly. In the unitary gauge, the fields ($H$ and $S$) can be parameterized as,
\begin{equation}
  H =
  \begin{pmatrix}
    0
    \\
    \frac{h}{\sqrt{2}}
  \end{pmatrix},
  \qquad
  S = \frac{s + i \chi}{2},
\end{equation}

In this paper, we only focus on the single-field inflation as one of many possibilities, where we set the global vacuum as $(v_h, v_s)$ with $v_{h} = 246.2~\mathrm{GeV}$ being the $H$ VEV and $v_s$ being the VEV of $S$ real part. They spontaneously break the electroweak symmetry and the $\mathbb{Z}_{3}$ symmetry after the end of the inflation.
We find that the imaginary part of $S$, the pseudo-Goldstone $\chi$, is still stable.
So the real and imaginary parts of the complex scalar field can be the inflaton and the dark matter respectively.
There exists only the inflaton in the inflation stage; the dark matter and Higgs particles are produced until the reheating stage, by the inflaton decays.


\section{Inflation in the metric and Palatini formalisms}\label{sec:Inf}

In the inflation stage, there's only the inflation field, thus the action can be simplified as
\begin{align}
S_J = \int d^{4}x \sqrt{-g_{J}} \left[ - \frac{M_{pl}^{2}+\beta s^2}{2} R+ \partial_{\mu}s \partial^\mu s^\dagger - V(s) \right],
\end{align}
where the scalar potential $V(s)=\frac{\lambda_{S}}{4}s^{4} + \frac{\mu_3}{2 \sqrt{2}}s^{3} + \frac{\mu_s^{2}}{2}s^{2}$. The Ricci scalar $R$ can be handled in two ways. One way, the metric formalism, is that $R$ is solely determined by the metric $g_{\mu\nu}$. While the other way, Palatini formalism, is that $R$ is determined by both the metric $g_{\mu\nu}$ and the connection $\Gamma^\mu_{\nu\lambda}$. Note that the connection $\Gamma^\mu_{\nu\lambda}$ is taken to be independent of the metric. As the present of non-minimal coupling $\frac{-\beta}{2} s^2 R$, the two formalisms are will not equivalent and predict different Universe \cite{Bauer:2008zj}. To derive meaningful results under the usual processes of quantum field theory, we need to get rid of such non-minimal coupling, which can be implemented by the Weyl transformation,
\begin{align}
g_{E,\mu\nu} = \Omega^{2} g_{J,\mu\nu},\qquad {\rm with }\quad\Omega^{2} \equiv \frac{M_{pl}^{2} + \beta s^2}{M_{pl}^{2}},
\end{align}
the action is conveniently transformed into the canonical form of gravity in Einstein frame
\begin{align}
S_{E} = \int d^{4}x \sqrt{-g_{E}} \left[ - \frac{M_{pl}^{2}}{2}R_{E} + \frac{1}{2} \Pi(s) (\partial s)^{2} - U(s)\right],
\end{align}
where the potential $U(s)=\frac{V(s)}{\Omega^{4}}$, and the non-trivial kinetic term is given by
\begin{align}
\Pi(s)  \equiv \frac{1}{ \Omega^{2}}  +  \frac{\zeta}{ M_{pl}^{2}} \frac{6\beta^2 s^{2}}{\Omega^{4}},
&& \zeta = \begin{dcases}
1 & \text{(metric formalism) } \\
0 & \text{(Palatini formalism) }
\end{dcases}.
 \label{eq:2eq}
\end{align}

The kinetic term is easily canonicalized by $ \sqrt{\Pi(s)} \partial \phi = \partial s$, thus
\begin{align} \label{Eq:JordantoEinstein}
\frac{d \phi}{d s} = \sqrt{\Pi(s)}.
\end{align}

The kinematic properties during inflation will be characterized by two slow-roll conditions ($\epsilon(s)<<1$, and $\eta(s)<<1$). The slow-roll parameters are defined in the Einstein frame:
\begin{align}
\epsilon(s)
&\equiv \frac{M_{\rm Pl}^2}{2} \left(\frac{\partial U(s(\phi))/\partial \phi}{U(s)}\right)^2 = \frac{M_{\rm Pl}^2}{2\Pi(s)} \left(\frac{\partial U(s)/\partial s}{U(s)}\right)^2, \\
\eta(s) &\equiv M_{\rm Pl}^2 \frac{\partial^2 U(s)/\partial \phi^2}{U(s)} = \frac{M_{\rm Pl}^2}{\sqrt{\Pi(s)} U(s)} \frac{\partial}{\partial s} \left(\frac{1}{\sqrt{\Pi(s)}} \frac{\partial U(s)}{\partial s}\right),
\end{align}

As the universe evolves, the slow-roll conditions will be broken, and inflation will end. The corresponding value of inflation ($s_{e}$) is given via $\epsilon(s_{e}) =1$.
Then, the number of e-foldings during expansion from a pivot scale with $a_{k}$ to the end of inflation at $a_{e}$ is given by
\begin{align}
\label{Eq:exactefold}
N = \log \frac{a_e}{a_k} =\frac{1}{M_{pl}^2} \int_{\phi(s_e)}^{\phi(s_k)} d\phi  \frac{ U(s)  }{\partial U(s)/\partial \phi} = \int_{s_e}^{s_k}  \frac{ds}{M_{pl}} \sqrt{\frac{\Pi(s)}{2\epsilon(s)}},
\end{align}
where $s_k $ is the value of inflation that the pivot scale $k$ leaves the horizon. In this work, the pivot scale is chosen to be $k = 0.05~\text{Mpc}^{-1}$.

\section{Dark matter}\label{sec:DM}



A lot of analysis of DM phenomenology and neutrino physics were carried out in $Z_3$ models (e.g see \cite{Ma:2007gq,Belanger:2012zr, Bhattacharya:2017fid, Arcadi:2017vis,Cai:2018imb,Aoki:2014cja,Bonilla:2016diq,Ding:2016wbd,Saez:2022pwi} ).  There are extra new features of the $Z_3$ symmetric models than the ordinary $Z_2$ DM models. Such as there exist DM semi-annihilation processes where the annihilation
of two DM particles can give rise to a DM particle in the final state  ~\cite{Hambye:2008bq,DEramo:2010keq,Belanger:2012vp,Belanger:2014bga},  which play an important role in the freeze-out.
In this part, we will study the parameters of the tree-level potential using the physical quantities in the zero-temperature vacuum. The tree-level potential can be written as,
\begin{align}\label{eq:z3singlet}
V_0(h,s,\chi)&=\frac{\lambda_{H}}{4}h^{4}+\frac{\lambda_{S}}{4}s^{4}+\frac{\lambda_{S}}{4}\chi^{4}
+\frac{\lambda_{SH}}{4} h^{2}s^{2}+\frac{ \lambda_{SH}}{4} h^{2}\chi^{2}+\frac{\lambda_{S}}{2} s^{2} \chi^{2} \nonumber\\
&+\frac{\mu_3}{2 \sqrt{2}}s^{3}-\frac{3 \mu_3}{2 \sqrt{2}} s\chi^{2}+\frac{\mu_H^{2}}{2}h^{2}+\frac{\mu_s^{2}}{2}s^{2}+\frac{\mu_s^{2} }{2}\chi^{2}\;.
\end{align}

The potential needs to be stable, which will bring the global stationary point:
\begin{align}
\left.\frac{d V_{0}(h, s, \chi)}{d h}\right|_{h=v_h}=0\;,\left.\frac{d V_{0}(h, s, \chi)}{d s}\right|_{s=v_s}=0\;.
\end{align}

One can express the mass parameters as
\begin{align}
\mu_{H}^2=-\lambda_{H} v_h^2 - \frac{1}{2} \lambda_{SH} v_s^2\;,\mu_S^2=-\lambda_{S} v_s^2 - \frac{1}{2} \lambda_{SH} v_h^2 - \frac{3\sqrt{2}}{4} \mu_3 v_s \;.
\end{align}

The mixing mass matrix between Higgs and the real part of $S$ is given by,
\begin{equation}
  M^{2} =
  \begin{pmatrix}
    2 \lambda_{H} v_h^{2} & \lambda_{SH} v_h v_{s}
    \\
    \lambda_{SH} v_h v_{s} & 2 \lambda_{S} v_{s}^{2} + \frac{3}{2 \sqrt{2}} \mu_{3} v_{s}
  \end{pmatrix}.
\label{eq:mass:matrix}
\end{equation}

To eliminate those mixing and get two mass eigenvalue $m_{h_1}$ (the SM Higgs boson mass) and $m_{h_2}$, one needs to introduce a orthogonal rotation matrix
\begin{equation}
  R =
  \begin{pmatrix}
    \cos \theta & -\sin \theta
    \\
    \sin \theta & \cos \theta
  \end{pmatrix}
\end{equation}

The expression of $m_{h_1}$ and $m_{h_2}$ can be obtained via $\diag (m_{h_1}^{2}, m_{h_2}^{2}) = R^{T} M^{2} R$, as follows:
\begin{align}
\label{eq:}
&m_{h_1}^2 = \frac{1}{4} v_s (8 v_s \lambda_{S}+3 \sqrt{2} \mu_3) \sin\theta^{2} +2 v_h \cos\theta(v \lambda_{H} \cos\theta+v_s \lambda_{SH} \sin\theta)\;, \nonumber\\
&m_{h_2}^2 = \frac{1}{4} v_s (8 v_s \lambda_{S}+3 \sqrt{2} \mu_3) \cos\theta^{2} - 2 v_h v_s \lambda_{SH} \cos\theta \sin\theta+2 v_h^{2} \lambda_{H} \sin\theta^{2}\;,
\end{align}
where the mixing angle $\theta$ should satisfy the following relationship,
\begin{equation}
   \tan 2 \theta = \frac{\lambda_{SH}  v_h v_{s}}{\lambda_{H} v_h^{2} - \lambda_{S} v_{s}^{2} - \frac{3}{4 \sqrt{2}} \mu_{3} v_{s}}.
\end{equation}

In the vacuum $(v_h, v_s)$, the mass of the pseudoscalar $\chi$ can be written as
\begin{equation}
  m_{\chi}^{2} = - \frac{9}{2 \sqrt{2}} \mu_{3} v_{s}.
\end{equation}

Now we can express the potential parameters in terms of physical quantities, i.e., the masses $m_{h_1}$ and $m_{h_2}$, the mixing angle $\theta$, the pseudoscalar mass $m_{\chi}$, and the VEVs $v_{h}$ and $v_{s}$.

\begin{itemize}
  \item For the coupling coefficients and $\mu_3$, their expressions can be obtained via the mass diagonalization condition and the mass $m_{\chi}^{2}$, as follows:
  \begin{align}
  \lambda_{H} &= \frac{ m_{h_1}^{2} + m_{h_2}^{2} + (m_{h_1}^{2} - m_{h_2}^{2}) \cos 2 \theta }{4 v_{h}^{2}},\label{Eq:lh}
  \\
  \lambda_{S} &= \frac{3(m_{h_1}^{2} + m_{h_2}^{2}) + 2 m_{\chi}^{2} + 3 (m_{h_2}^{2} - m_{h_1}^{2}) \cos 2 \theta}{12 v_{s}^{2}},\label{Eq:ls}
  \\
  \lambda_{SH} &= \frac{(m_{h_1}^{2} - m_{h_2}^{2}) \sin 2 \theta}{2 v_{s} v_{h}},\label{Eq:lsh}
  \\
  \mu_{3} &= -\frac{2 \sqrt{2}}{9} \frac{m_{\chi}^{2}}{v_{s}}.
  \end{align}

  \item For the coefficients of the mass terms, the expressions can be obtained via substitution of the coupling coefficients and $\mu_3$, as follows:
  \begin{align}
  \mu_{H}^{2} &= -\frac{1}{4} (m_{h_1}^{2} + m_{h_2}^{2}) + \frac{1}{4 v_{h}} (m_{h_2}^{2} - m_{h_1}^{2})
  \notag
  \\
  & \times (v_{h} \cos 2 \theta + v_{s} \sin 2 \theta),
  \\
  \mu_{S}^{2} &= -\frac{1}{4} (m_{h_1}^{2} + m_{h_2}^{2}) + \frac{1}{6} m_{\chi}^{2} + \frac{1}{4 v_{s}}
  (m_{h_1}^{2} - m_{h_2}^{2})
  \notag
  \\
  & \times (v_{s} \cos 2 \theta - v_{h} \sin 2 \theta).
\end{align}
\end{itemize}

The DM annihilation processes depend on the mass of DM $\chi$,  and the coupling between DM and other particles. The dominant DM pair annihilation channels include: $\chi \chi \to (h_{2}h_{2}, h_{1}h_{1}, t\bar{t},h_{1}h_{2}, W^+W^-, ZZ,b\bar{b})$. In this paper, we calculate the relic density of $\chi$ by using the package micrOMEGAs \cite{Belanger:2018ccd}.
For a heavy new scalar $h_2$ ($m_{h_2}>m_{\chi}$), the dominant DM annihilation processes are into Higgs pairs $h_1 h_1$, top quark pairs $t\Bar{t}$ and scalar pairs $h_1 h_2$.

For the mixing angle $\theta$ between $h$ and $s$, it is constrained from the measurements of the Higgs couplings at the LHC to $\abs{\sin \theta} \leq 0.37$ for $m_{h_2} \gtrsim m_{h_1}$ and $\abs{\sin \theta} \leq 0.5$ for $m_{h_2} \lesssim m_{h_1}$~\cite{Ilnicka:2018def}.
The dark matter relic density $\Omega h^{2} = 0.12 \pm 0.0012$, taken from recent Planck data~\cite{Planck:2018vyg}.

\section{Constraints}\label{sec:Con}

We impose various theoretical and experimental constraints on the parameter spaces.

\subsection{Theoretical constraints}

\begin{itemize}
  \item The potential should have the minimum value to ensure the stability of the  potential, which leads to the following constraints,
  \begin{equation}
  \lambda_{H} > 0, \quad \lambda_{S} > 0, \quad \lambda_{ SH} + 2 \sqrt{\lambda_{H}} \lambda_{ S} > 0.
  \end{equation}
  \item To keep this model in the perturbation computability, the couplings need to meet the following constraints~\cite{Lerner:2009xg,Kannike:2019mzk},
  \begin{equation}
  \abs{\lambda_{H}} \leqslant \frac{2}{3} \pi, \quad \abs{\lambda_{S}} \leqslant \pi, \quad \abs{\lambda_{SH}} \leqslant 4 \pi.
  \end{equation}
  \item  We force the couplings to be unitary by requiring that the    couplings of the potential meet the following constraints~\cite{Kannike:2019mzk},
  \begin{align}
  \abs{\lambda_{H}} \leqslant 4 \pi,  \quad\abs{\lambda_{S}} \leqslant 4 \pi, \quad  \abs{\lambda_{SH}} &\leqslant 8 \pi,
  \\
  \lvert 3 \lambda_{H} + 2 \lambda_{S}
   \pm \sqrt{9 \lambda_{H}^2 - 12 \lambda_{H} \lambda_{S} + 4 \lambda_{S}^2 + 2 \lambda_{SH}^{2}} \rvert &\leqslant 8 \pi,
  \end{align}
  \item Assume that the global vacuum point of the potential is $(v_h, v_s)$, which would impose a mass-dependent constraint, as follows,
  \begin{equation}
  \label{Eq:global}
  m_{\chi}^{2} < \frac{9 m_{h_1}^{2} m_{h_2}^{2}}{m_{h_1}^{2} \cos^2 \theta + m_{h_2}^{2} \sin^{2} \theta}.
  \end{equation}
\end{itemize}

\subsection{CMB Constraints}

With the value of inflation $(s_k)$, one can calculate the tensor-to-scalar ratio ($r$) and the scalar spectral index ($n_s$) as,
\begin{eqnarray}
r &=&16\,\epsilon(s_k), \\
n_s &=&1-6\epsilon(s_k) +2\eta(s_k).
\end{eqnarray}
The latest Planck constraints on $r$ and $n_s$ can be found in Fig.~\ref{Fig:ns_r},  with lighter and darker green regions being the  $1\sigma$ and $2\sigma$ experiment errors of BAO, BICEP/KECK, and Planck data respectively~\cite{BICEP:2021xfz}.

Noting that inflation occurs at very high energy scale, we use RGE to evaluate the coupling coefficients from the electroweak scale to the inflation scale. The beta functions of the scalar quartic couplings are given by
\begin{eqnarray}
\beta_ {\lambda_{H}}&=& \frac{3 g_1^4}{128 \pi ^2}+\frac{3g_{1}^2 g_{2}^2}{64 \pi ^2}-\frac{3 g_{1}^2\lambda_{H}}{16 \pi ^2}+\frac{9 g_{2}^4}{128 \pi ^2}-\frac{9g_{2}^2\lambda_{H}}{16 \pi ^2}-\frac{3 g_{t}^4}{8 \pi ^2}\nonumber\\
&&+\frac{3 g_{t}^2\lambda_{H}}{4 \pi ^2}+\frac{3 \lambda_{H}^2}{2 \pi ^2}+\frac{\lambda_{SH}^2}{16 \pi ^2}\; ,
\\
\beta_ {\lambda_{S}}&=& \frac{\lambda_{SH}^2}{8 \pi ^2}+\frac{5 \lambda_{S}^2}{4 \pi ^2}\; ,
\\
\beta_{ \lambda_{SH}}&=& -\frac{3 g_{1}^2\lambda_{SH}}{32 \pi ^2}-\frac{9 g_{2}^2 \lambda_{SH}}{32 \pi ^2}+\frac{3g_{t}^2\lambda_{SH}}{8 \pi ^2}+\frac{3\lambda_{H} \lambda_{SH}}{4 \pi ^2}+\frac{\lambda_{SH}^2}{4 \pi ^2}+\frac{\lambda_{SH}\lambda_{S}}{2 \pi ^2}\; ,\nonumber\\
\end{eqnarray}
where the initial conditions of the RGEs can be obtained from Ref.\cite{Buttazzo:2013uya}.
The running of the gauge couplings and the top quark Yukawa coupling can refer to~\cite{Ford:1992mv}.

\section{Numerical analysis}\label{sec:Num}

The usual strategy of numerical analysis is to directly take the physical masses and $\theta$ as input parameters to analyze the physical phenomenon.
However, part of the physical information is inevitably missed due to non-systematic considerations in the parameter space.
For a more comprehensive analysis, the couplings $\lambda_{H}, \lambda_{S}, \lambda_{SH}$, and $\theta$ are considered as the input parameters on them. The values of physical quantities, such as the DM mass, can be obtained by their relations with the inputs.

\begin{figure}[htbp]
\centering
\includegraphics[width=0.325\textwidth]{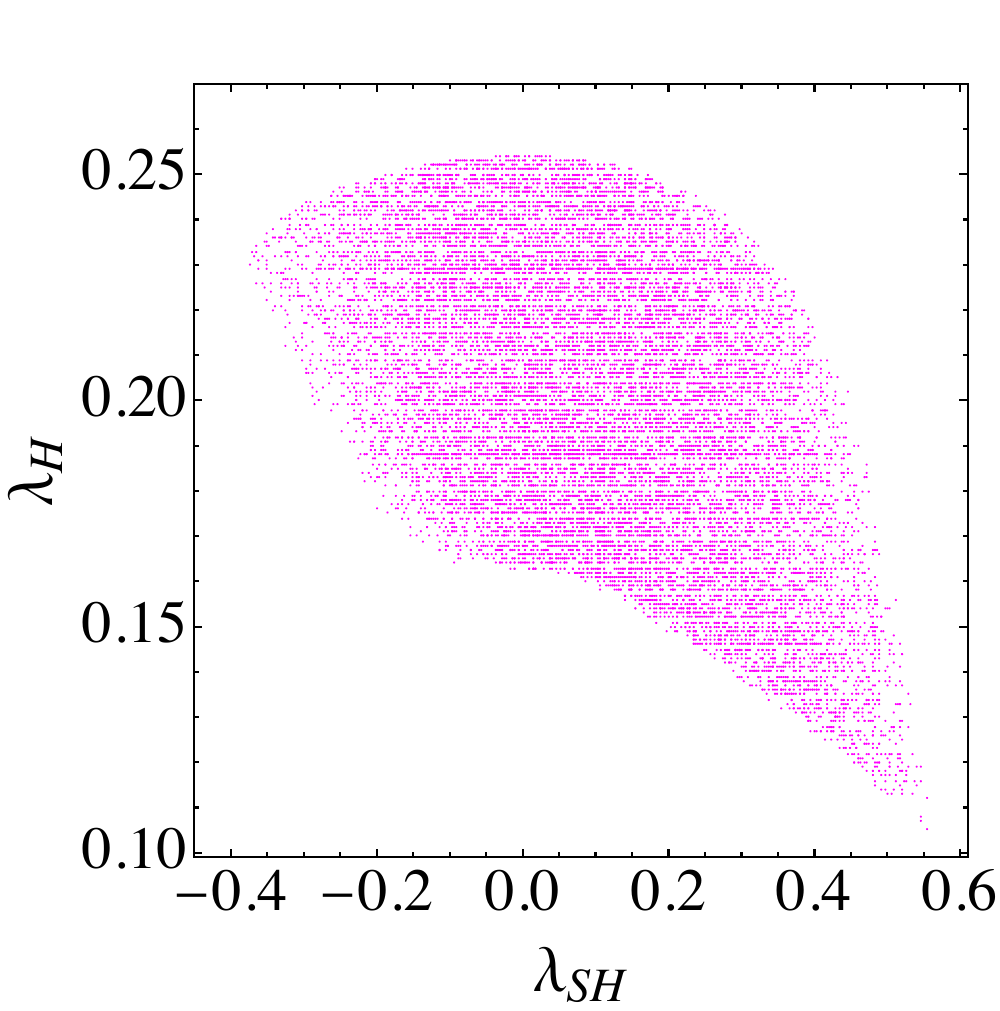}
\includegraphics[width=0.326\textwidth]{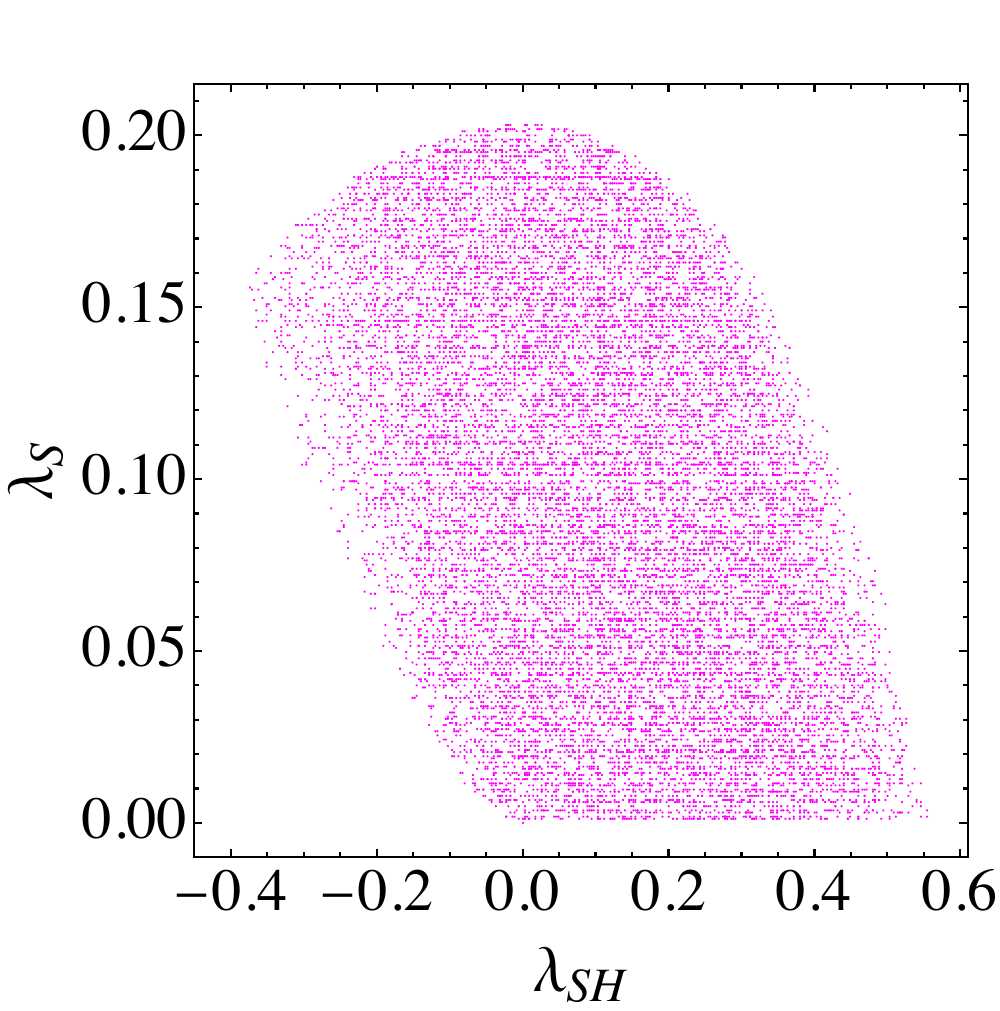}
\includegraphics[width=0.316\textwidth]{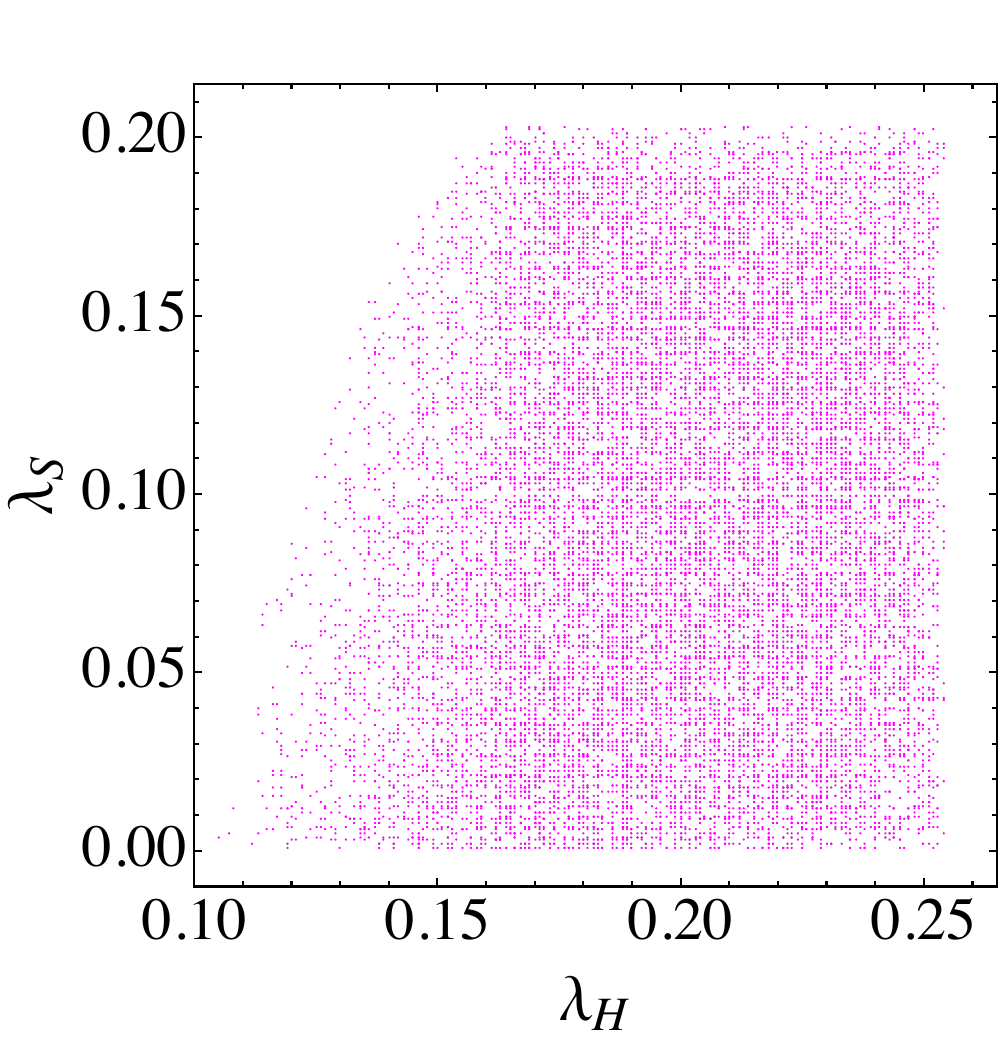}
\caption{Feasible points for couplings $\lambda_H, \lambda_S, \lambda_{SH}$, constrained by Perturbativity + Unitarity + Stability from the electroweak scale to Planck scale.}
\label{Fig:3setcoupling}
\end{figure}

We first randomly take the value of the couplings within the ranges of $\lambda_{H} \in (0, 1]$, $\lambda_{S} \in (0, 1]$ and $\lambda_{SH} \in [-1, 1]$, and impose the theoretical constraints on them. The feasible points are shown in the three different panels of Fig.~\ref{Fig:3setcoupling}.
The specific values for each coupling set can be obtained by combining the three diagrams simultaneously.
It is worth noting that these coupling points satisfy theoretical constraints from the electroweak scale to the Planck scale.

\begin{figure}[htbp]
\centering
\includegraphics[width=0.55\textwidth]{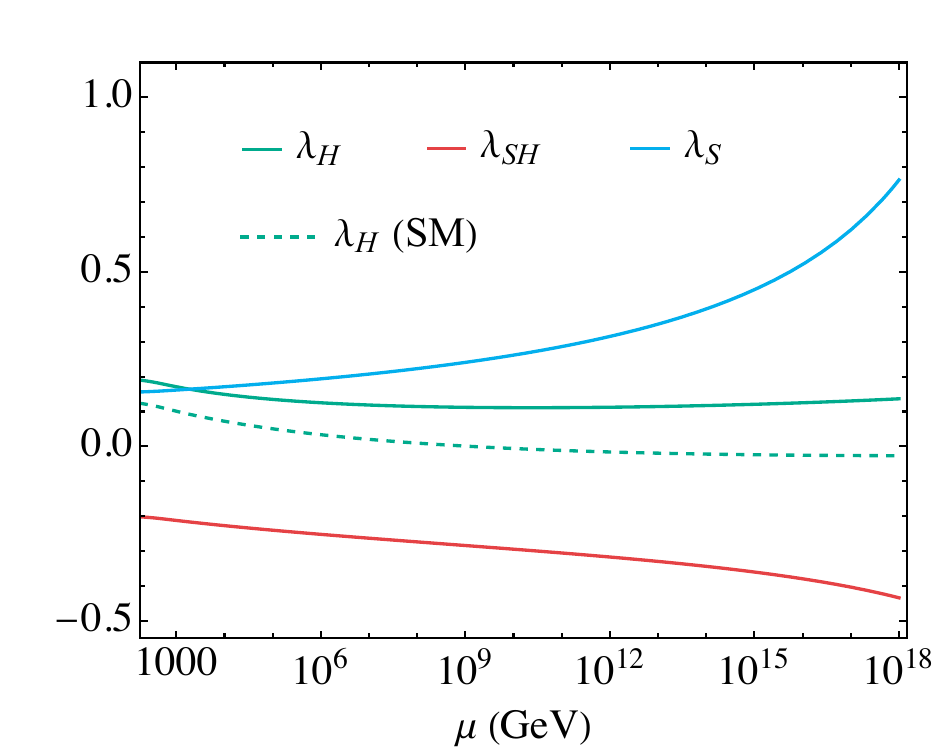}
\caption{The running behavior of couplings $\lambda_{H}$, $\lambda_{SH}$, and $\lambda_{S}$ with benchmark point {$\lambda_{{H}}=0.196$, $\lambda_{SH}=-0.199$, and $\lambda_{{S}}=0.155$}. The dashed green line stands for the running coupling $\lambda_H$ in SM, for a comparison.}
\label{fig:Example}
\end{figure}

As an example, we select a point set {$\lambda_{H}=0.196$, $\lambda_{SH}=-0.199$, and $\lambda_{S}=0.155$}, to study their running relation with the energy scales in Fig.~\ref{fig:Example}.
For comparison, the running of the Higgs quartic coupling in SM is shown as green dotted line.
As the energy scale increases, $\lambda_{H}$ decreases monotonically. When $\mu \gtrsim 10^{10}~\rm{GeV}$, $\lambda_{H}$ becomes negative in SM, while $\lambda_{H}>0$ in this model. It means that the vacuum stability of the Higgs potential will be guaranteed with the new degree of freedoms.

\begin{figure}[htbp]
\centering
\includegraphics[width=0.52\textwidth]{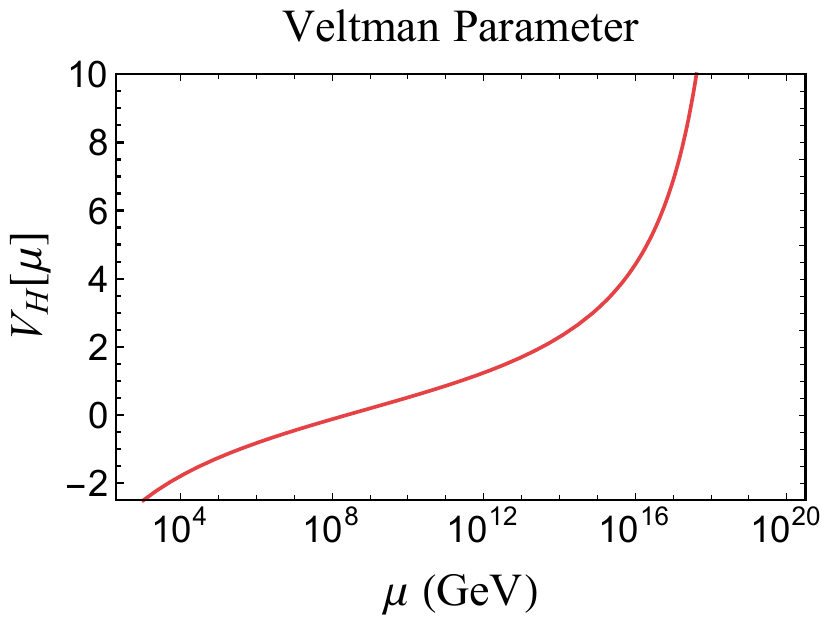}
\caption{The running Higgs Vetman parameter.}
\label{Fig:VP}
\end{figure}

In our scalar extended model, the corrected unphysical bare higgs mass could be expressed as
follows:
\begin{align}
m_h^2&={m^{bare}_h}^2 +\delta m_h^2\;.
\end{align}
where
\begin{align}
\delta m_h^2&=\frac{\Lambda^2}{16\pi^2}(6\lambda_H+\frac{3}{4}g_1^{2}+\frac{9}{4}g_2^{2}
-6y_t^2+\lambda_{SH})\;.
\end{align}
Both the real and imaginary of the scalar $S$ couple to higgs, which will lead to a loop correction for the higgs mass, and the total correction can be expressed as $\frac{\Lambda^2}{16\pi^2}{\lambda_{SH}}$.
We use the
same benchmark point to study the fine-tuning issue. As pointed out in Refs.~\cite{Baer:2013gva,Chakraborty:2014oma}, we can say that the Higgs mass is natural and has not to fine-tuned if $|\delta m_h^2|\leq m_h^2$.  For this set of coupling coefficients, one can obtain the cut-off scale $\Lambda \sim 824~\rm{GeV}$, which means our model is valid up to near TeV scale. For strict Veltman Condition $\delta m_h^2=0$, which implies the Veltman Parameter $V_H[\mu]=(6\lambda_H+\frac{3}{4}g_1^{2}+\frac{9}{4}g_2^{2}-6y_t^2+\lambda_{SH})=0$. The Veltman Parameter $V_H[\mu]$ as a function of energy scale is shown in the following Fig.~\ref{Fig:VP}, which imply that $V_H[\mu]=0$ with $\mu\approx10^8~\rm{GeV}$, and this energy scale is about $11$ order of magnitude smaller than that of SM~\cite{Habibolahi:2022rcd}, which would release the fine-tuning problem.

\begin{figure}[htbp]
\centering
\includegraphics[width=0.62\textwidth]{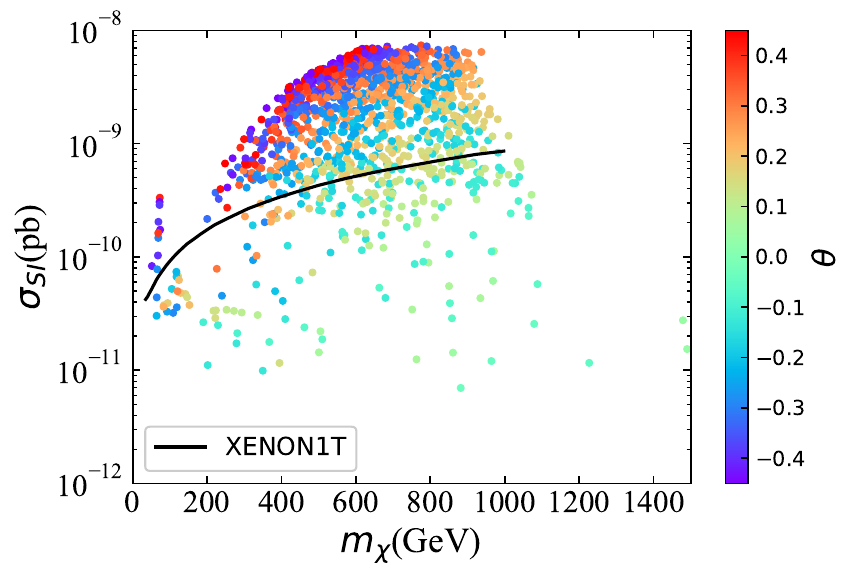}
\caption{Parameter points further constrained by DM relic density and DM direct detection limits. The colored points, which change with different mixing angle $\theta$,  satisfy the correct relic density.
The black line is the upper bound from the XENON1T experiment.}
\label{Fig:dir_dete}
\end{figure}

After further considering the limit from the global vacuum (Eq.~\ref{Eq:global}) and the DM relic density~\cite{Planck:2018vyg}, the remaining feasible points are shown in Fig.~\ref{Fig:dir_dete}. The color depth of the points corresponds to the size of the $\theta$ value. The black line stands for the DM direct detection limits from the latest XENON1T results ~\cite{XENON:2018voc}.
We found that the darker points with larger $|\theta|$ values are excluded, because the stronger the mixing between $h$ and $s$, i.e., the greater the $|\theta|$, the easier to be detected. To this end, the range of the mixing angle  $\theta$ can be determined.

\begin{figure}[htbp]
\centering
\includegraphics[width=0.45\textwidth]{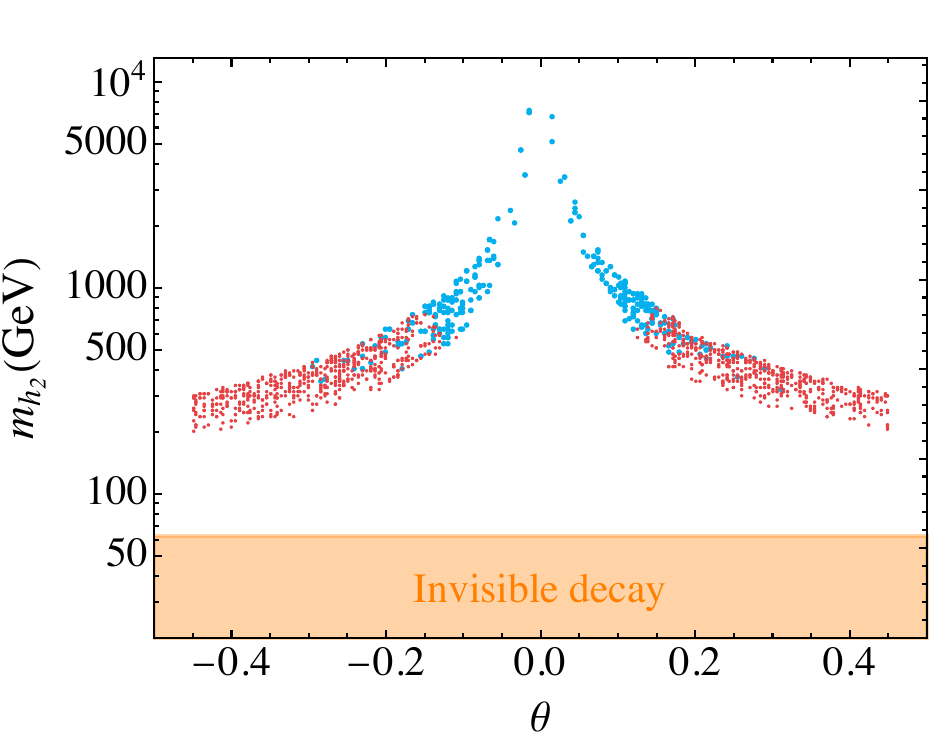}
\includegraphics[width=0.45\textwidth]{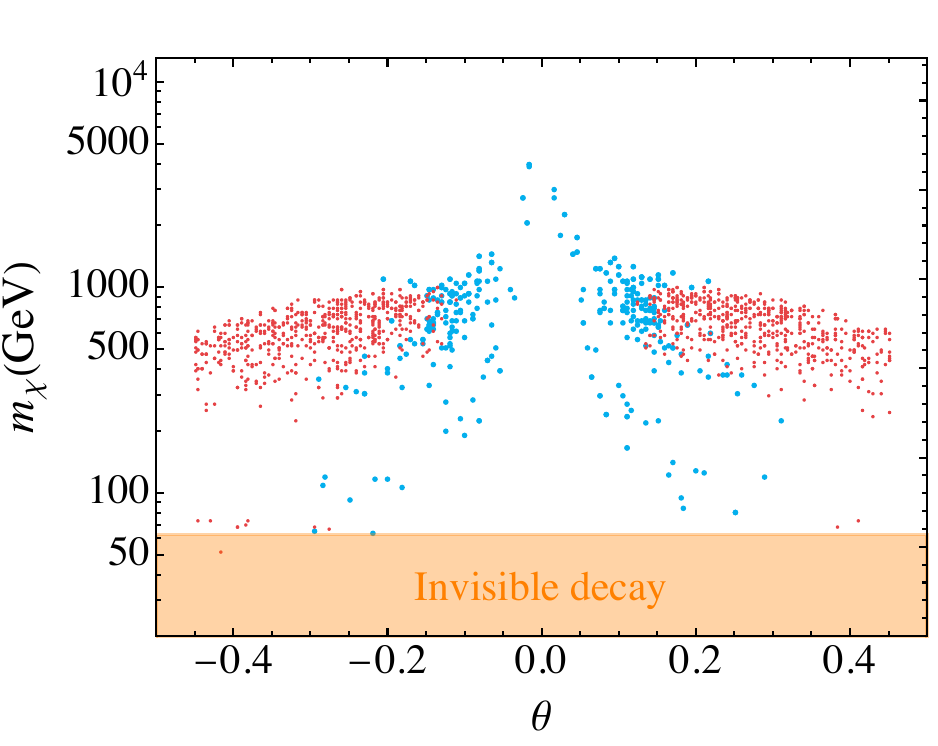}
\caption{Scalar masses as functions of the mixing angle $\theta$.
The red points are the exclusion points by XENON1T experiment, and the blue points are the remaining feasible points. Note that above the orange region there is not contribution to the Higgs invisible decay.}
\label{Fig:mass_theta}
\end{figure}

To make the relationship between $\theta$ and scalar masses more clear, we transform the points in Fig.~\ref{Fig:dir_dete} into the ($\theta, m_{h_2}$) and ($\theta, m_\chi$) planes, as shown in Fig.~\ref{Fig:mass_theta}.
The red points are the exclusion points by XENON1T experiment, and the blue points are the remaining feasible points. The corresponding feasible couplings are $\lambda_H\sim[0.161,0.252]$, $\lambda_S\sim[0.007,0.196]$, $\lambda_{SH}\sim[-0.254,0.232]$.
Both plots in Fig.~\ref{Fig:mass_theta} imply that the mixing angle $|\theta| \lesssim  0.25$, which satisfies the measurements of the Higgs couplings at the LHC, i.e., $\abs{\sin \theta} \leq 0.37$ for $m_{h_2} \gtrsim m_{h_1}$, and $\abs{\sin \theta} \leq 0.5$ for $m_{h_2} \lesssim m_{h_1}$~\cite{Ilnicka:2018def}.
We also find that $m_{h_2} \gtrsim 300~\rm{GeV}$ and DM mass $m_\chi \gtrsim 80 ~\rm{GeV}$, which means none of them contributes to Higgs invisible decay.

\begin{figure}[htbp]
\centering
\includegraphics[width=0.55\textwidth]{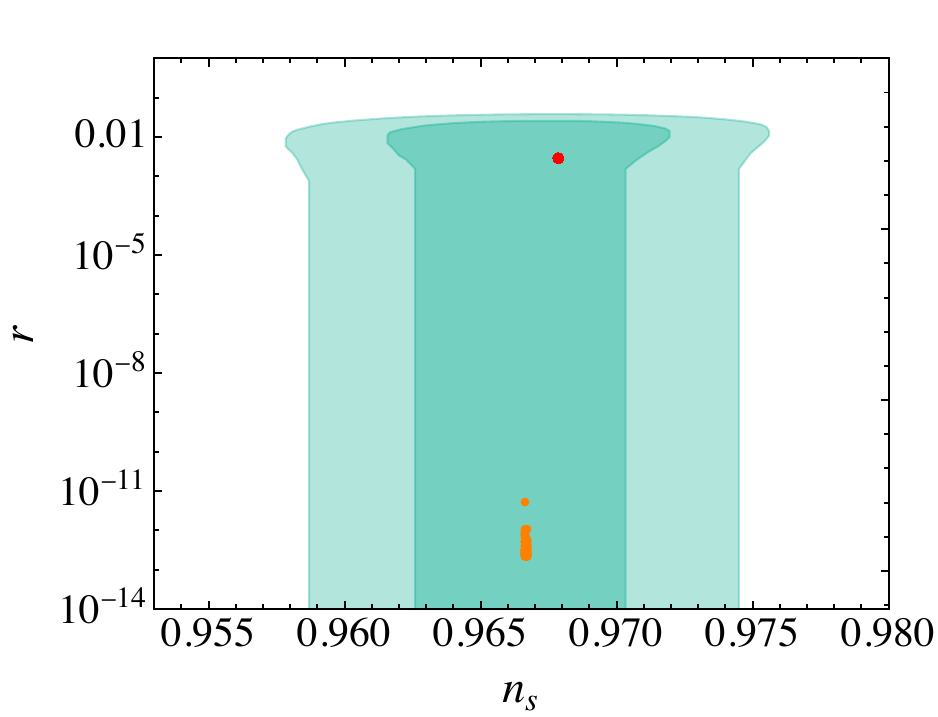}
\caption{The observables $r$ and $n_s$. The latest combination $1\sigma$ and $2\sigma$ experiment errors of BAO, BICEP/KECK, and Planck data are shown in the darker and lighter green regions, respectively \cite{BICEP:2021xfz}. The red points correspond to the metric formalism ($\zeta=1$), while the orange points correspond to Palatini formalism ($\zeta=0$).}
\label{Fig:ns_r}
\end{figure}

Finally, we study the inflation in the remaining parameter space. Fig.~\ref{Fig:ns_r} indicates the relationship between $n_s$ and $r$.
The red points correspond to the metric formalism ($\zeta=1$), while the orange points correspond to Palatini formalism ($\zeta=0$). The latest combination $1\sigma$ and $2\sigma$ experiment errors of BAO, BICEP/KECK, and Planck data are shown in the darker and lighter green regions, respectively \cite{BICEP:2021xfz}. For the metric formalism scenario, the corresponding non-minimal coupling is $\beta \sim 10^4$. The red points are almost completely overlapped, which indicates that both $n_s$ and $r$ are insensitive to the changes in the quartic coupling $\lambda_S$. For Palatini scenario, the corresponding non-minimal coupling is $\beta \sim 10^9$.
The orange points form a straight vertical line, which indicates that the change of $\lambda_S$ has almost no effect on $n_s$ and a small effect on $r$. However, since the value of $r$ is too small, this effect does not affect the experimental observations.
With the large hierarchy in $r$, the inflation under the two forms can be distinguished by measuring the tensor-to-scalar ratio.

\section{Summary}\label{sec:Sum}

In this paper, two important aspects of modern physics, DM and inflation, are jointly studied by employing a complex scalar model with $\mathbb{Z}_{3}$ symmetry. The two topics correspond to the low-energy and high-energy scale physics, which can be properly connected by RGE evolution. At the same time, non-minimal coupling corrections handled in two ways (the metric and Palatini formalism), are introduced to realize the slow-roll inflation.

We test the parameter space by considering all the theoretical constraints, and the results can refer to Fig.~\ref{Fig:3setcoupling}. Also, the couplings are required to satisfy the constraints from the low energy scale to the inflation scale, which eliminates most of the parameter space.
After considering the mass limits from the global vacuum requirement, the relics density of DM from the latest observation, and the upper bounds on DM-nucleon scattering, we obtain the mixing angle $|\theta| \lesssim 0.25$, which is in agreement with the LHC experimental constraints. Also the corresponding scalar masses result in $m_\chi \gtrsim 80~\rm{GeV}$ and $m_{h_2} \gtrsim 300~\rm{GeV}$.

For the slow-rolling inflation, compared with the quadratic and cubic inflation term, the quartic term with coupling $\lambda_{S}$ dominates the behavior of potential. As $\lambda_{S} \sim O(0.1)$, so that $\lambda_{S}$ has little effect on inflation. Numerical analysis shows that both slow-rolling inflation in the forms of metric and Palatini formalisms can be realized, as which is consistent with the cosmological constraints on $n_s$ and $r$. More specifically, the magnitudes of the non-minimal coupling coefficient $\beta$ in the form of metric and Palatini formulae are $\sim 10^4$ and $\sim 10^9$, respectively. It suggests that the non-minimal coupling strength has a huge impact on inflation. For the inflation observations in the metric and Palatini formalisms, the scalar spectral indices $(n_s)$ are both $\sim 0.965$, and the tensor-to-scalar ratio are $\sim 10^{-3}$ and $\lesssim 10^{-11}$, which indicates that the inflation in the metric formalism may be tested by the future experiments with higher precision.
\hspace{2cm}

\acknowledgments

Wei Cheng was supported by Chongqing Natural Science Foundation project under Grant No. CSTB2022NSCQ-MSX0432, by Science and Technology Research Project of Chongqing Education Commission under Grant No. KJQN202200621, and by Chongqing Human Resources and Social Security Administration Program under Grants No. D63012022005.
Xuewen Liu was supported by the National Natural Science Foundation of China under Grants No.12005180, by the Natural Science Foundation of Shandong Province under Grants No.ZR2020QA083, and by the Project of Shandong Province Higher Educational Science and Technology Program under Grants No. 2019KJJ007.
Ruiyu Zhou was supported by Chongqing Natural Science Foundation project under Grant No. CSTB2022NSCQ-MSX0534.
This work was also supported in part by the Fundamental Research Funds for the Central Universities under Grant No. 2021CDJQY-011,
and by the National Natural Science Foundation of China under Grant  No. 12147102.

\bibliography{lit}

\end{document}